\documentclass[11pt]{article}

\usepackage{amsfonts}
\usepackage{epsf}
\usepackage{graphicx}
\usepackage{psfrag}

\newcommand{\rme}{\mathrm{e}}
\newcommand{\rmi}{\mathrm{i}}
\newcommand{\rmd}{\mathrm{d}}
\renewcommand{\qquad}{\hspace*{25pt}}
\newcommand{\Tr}{\mathop{\mathrm{Tr}}\nolimits}

\newcommand{\Or}{\mathord{\mathrm{O}}} 

\begin{document}

\begin{center}
{\Large{\bf Two-parametric hyperbolic octagons and reduced Teichm\"uller space in genus two}}

\vspace*{0.5cm}
{A.V. Nazarenko}

\vspace*{0.5cm}
{Bogolyubov Institute for Theoretical Physics,\\
14-b, Metrologichna Str., Kiev 03680, Ukraine\\
nazarenko@bitp.kiev.ua}
\end{center}

\begin{abstract}
It is explored a model of compact Riemann surfaces in genus two, represented
geometrically by two-parametric hyperbolic octagons with an order four automorphism.
We compute the generators of associated isometry group and give a real-analytic
description of corresponding Teichm\"uller space, parametrized by the Fenchel-Nielsen
variables, in terms of geometric data. We state the structure of parameter space
by computing the Weil-Petersson symplectic two-form and analyzing the isoperimetric orbits.
The results of the paper may be interesting due to their applications to the quantum geometry,
chaotic systems and low-dimensional gravity.
\end{abstract}

\section{Introduction}

The Riemann surface of genus two serves as geometry carrier in a great number of the
models\footnote{Here we refer to few works but directly related to a given topic.}
of string theory~\cite{DHP}, statistical physics~\cite{Hurt,Naz}, chaology~\cite{Gutz,N,N1},
low-dimensional gravity~\cite{NR,Lo}. Problems, in which surface geometry is not fixed
and is developing in time, have a special interest. The changes of underlying geometry
can be described in different ways, for instance, by evolution equations, by averaging
over surface moduli or parameters, etc. Then it is naturally to require the surface
deformation to be represented by continuous and smooth trajectory in a some
space with the properties which should be carefully studied.

Although the case of genus two gives us access to quite explicit calculations, most
of problems cannot be solved in general. This fact forces us to concentrate on
a family of the surfaces with the reduced number of geometric degrees of freedom.
Using the more convenient geometric approach we consider the surfaces represented
by two-parametric hyperbolic octagons embedded into the unit disk.

Assuming that an octagon form remains the same under rotation by $\pi/2$, we first
construct the fundamental domain with opposite sides identified and the associated
Fuchsian group, using as ``input'' the two real parameters: length and angle,
determining the position of vertices, i.e. the octagon geometry. Although the
general formalism linking the geometric data and the Fuchsian group is
known~\cite{ABCKS}, we pay great attention to manifest dependence of octagon boundary
segments and isometry group generators on these parameters in order to make the
functions straightforwardly applicable in forthcoming calculations.

We aim to investigate a real-analytic structure of parameter space, that is dictated
by isometry group of Teichm\"uller space, usually called as the mapping class group,
and essentially determines an initial octagon evolution in various physical problems.
Thus, to realise this, we introduce Teichm\"uller space for the surfaces under
consideration as a subset of total Teichm\"uller space for all surfaces in genus two
and compute the Fenchel-Nielsen variables regarding as the global coordinates on it.

We perform main analysis (in Section 3) within the Weil-Petersson geometry allowing
us to endow the parameter space with the symplectic two-form which is invariant by
definition under action of the mapping class group. Key tool is the Wolpert's
formula~\cite{Wo85} permitting us to express this form in terms of the Fenchel-Nielsen
variables. As the result, we shall see that the accessibility domain of
geometric data used is the symplectic orbifold. Furthermore, the symmetry group of
the reduced Teichm\"uller space is expected to be wide than the mapping class group
because of geometric constraints imposed. Note that the involution of the surfaces
with an order four automorphism and the associated generators are diskussed in
details in \cite{Sil}.

We supplement our results by description of the isoperimetric orbits in the
parameter space (Section 4), what gives us additional information about the
structure of this space and reflects a particular diffeomorphism of octagon.
On the other hand, the dense set of isoperimetric orbits can serve as a tool
for further geometric quantization, independent of the octagon automorphisms
and pants decompositions. Physically, such an approach leads to a study of the
quantum spectra and quantum fluctuations of geometric quantities of
the objects with the same topology/geometry but of different nature.

\section{Model octagons and their symmetries}

We will deal with the Poincar\'e model of two-dimensional hyperbolic space, namely,
with the open unit disk centered at the origin,
\begin{equation}
\mathbb{D}=\left\{z=x+\rmi y\in\mathbb{C}||z|<1\right\},
\end{equation}
endowed with the metric
\begin{equation}\label{Poin}
\rmd s^2=4\frac{\rmd x^2+\rmd y^2}{(1-x^2-y^2)^2}
\end{equation}
of the Gaussian curvature $K=-1$.

Hyperbolic distance between complex coordinates $z$ and $w$ in space $(\mathbb{D},\rmd s^2)$
is denoted by ${\rm dist}_{\mathbb{D}}(z,w)$ and determined from relation:
\begin{equation}
\cosh{{\rm dist}_{\mathbb{D}}(z,w)}=1+\frac{2|z-w|^2}{(1-|z|^2)(1-|w|^2)},
\end{equation}
where $|z-w|$ is the Euclidean distance.

Solution to the geodesic equation in $(\mathbb{D},\rmd s^2)$ is the function:
\begin{equation}\label{xy}
z(s)=\frac{\cosh{s}+\rmi R\sinh{s}}{\sqrt{1+R^2}\cosh{s}+R}\exp{(-\rmi\phi)},
\end{equation}
which is defined in the interval $s\in(-\infty,+\infty)$ and describes an arc
inside $\mathbb{D}$ with radius $R$ and center at the point
$z_0=\sqrt{1+R^2}\exp{(\rmi\phi)}$ lying beyond the unit disk. In particular
case, the geodesics emanating from the origin are the Euclidean straight lines
(diameters). All geodesics intersect the boundary $\partial\mathbb{D}$ orthogonally.

The group of all orientation-preserving isometries of $(\mathbb{D},\rmd s^2)$,
denoted by ${\rm Isom^+(\mathbb{D})}$, acts via the M\"obius transformation:
\begin{equation}
z\mapsto \gamma[z]=\frac{uz+v}{\overline{v}z+\overline{u}},\quad
z\in\mathbb{D},
\end{equation}
where $u$ and $v$ are complex numbers satisfying relation $|u|^2-|v|^2=1$, and $\bar u$,
$\bar v$ are the complex conjugates. Thus, it is convenient to identify a generator
$\gamma$ with an element of group ${\rm SU(1,1)}/\{\pm1\}$,
\begin{equation}
{\rm SU(1,1)}=\left\{\left.
\left(\begin{array}{cc}
u&v\\
\overline{v}&\overline{u}
\end{array}\right)\right| u,v\in\mathbb{C},
|u|^2-|v|^2=1
\right\}.
\end{equation}

We shall concentrate on the properties of Riemann surface $S$ of genus $g=2$, which is
understood here as a compact two-dimensional orientable manifold with the Riemannian
metric of the constant negative curvature. Such a surface is obtained from a hyperbolic
simply connected octagon ${\cal F}$ in $\mathbb{D}$ via gluing opposite sides formed by
eight geodesic arcs, whose intersections serve as vertices.

In our model, which has been already declared and geometrically described in~\cite{Naz},
we assume that the vertices are at the points $a\exp{(\rmi k\pi/2)}$,
$b\exp{[\rmi(\alpha+k\pi/2)]}$, where $0<\alpha<\pi/2$, $0<a,b<1$, $k=\overline{0,3}$.
We also require the sum of the inner angles of ${\cal F}$ to be equal to $2\pi$. This is
the same as requiring ${\rm area}({\cal F})=2\pi(2g-2)=4\pi$ in accordance with the
Gauss-Bonnet theorem~\cite{DNF}.

Such an octagon (sketched in figure~1, left panel) is two-parametric: we choose a pair
$(a,\alpha)$ as independent real variables while parameter $b$ together with the
parameters of geodesics (sides) are functions of those. Although the surface of genus
$g$ is generally determined by $6g-6$ real parameters, we reduce 6-parametric object
to 2-parametric one by imposing constraints that simplifies considerations.
Therefore, the model octagon ${\cal F}$ can be viewed as a ``minimal deformation'' of
the regular hyperbolic octagon with $b=a=2^{-1/4}$, $\alpha=\pi/4$, well studied in
the context of the chaology (see, for example, \cite{N} and references therein).

The octagon we have obtained is stable under rotation by $\pi/2$. It means that the surface
has an order four automorphism. Note that the connection between this geometric model and
algebraic curves was intensively explored in the works of P.~Buser and R.~Silhol~(for example,
see \cite{Sil,BS} and references therein).

In this Section we review the connection between the model octagon ${\cal F}$, the
corresponding Riemann surface $S$ and the Fuchsian group 
$\Gamma\subset{\rm Isom^+(\mathbb{D})}$ isomorphic to fundamental
group $\pi_1(S)$ of surface. For more details we refer to \cite{ABCKS}.

\begin{figure}
\begin{center}
\includegraphics[width=4.5cm]{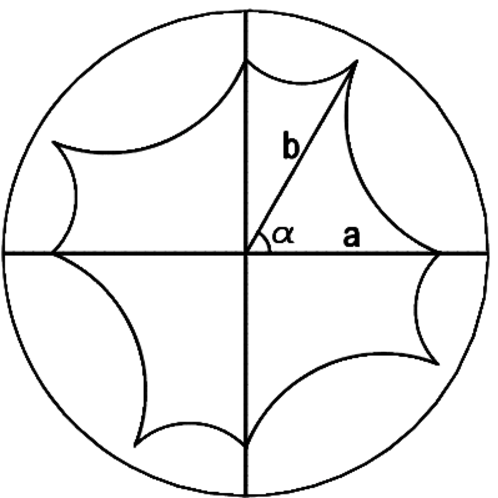}\qquad
\includegraphics[width=4.6cm]{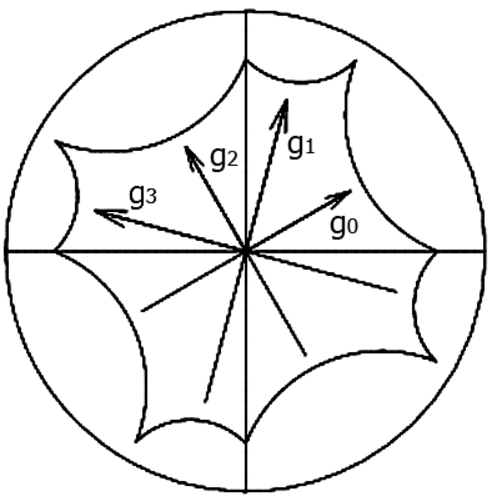}
\end{center}
\vspace*{-3mm}
\caption{\small Symmetric octagon with $a=0.8$, $\alpha=\pi/3$ and generators $g_k$ of
Fuchsian group.}
\end{figure}

In the case at a hand, the octagon boundary $\partial{\cal F}$ is formed by geodesics
of two kinds (labeled by ``$\pm$'' below). These geodesics are completely determined by the
radii $R_\pm$ and the angles $\phi_\pm+k\pi/2$, $k=\overline{0,3}$ (see (\ref{xy})),
defining the positions of the circle centers. Satisfying the conditions imposed above
(and collected in manifest form in Appendix of \cite{Naz}), we obtain that
\begin{equation}\label{R-phi}
R_\pm=\frac{1}{2a}\sqrt{T^2_\pm+(1-a^2)^2},\qquad
\phi_\pm=\arctan{\left[\left(\frac{T_\pm}{1+a^2}\right)^{\pm1}\right]},
\end{equation}
where $0<\phi_+<\alpha<\phi_-<\pi/2$, and
\begin{equation}
T_\pm=a^2\pm\tan{\tilde\alpha},
\qquad {\tilde\alpha}=\alpha-\pi/4.
\end{equation}

Moreover, introducing the inner angle $\beta$ by vertices $a\exp{(\rmi k\pi/2)}$
(the angle by vertices\\ $b\exp{[\rmi(\alpha+k\pi/2)]}$ is then equal to $\pi/2-\beta$)
such that
\begin{equation}
\tan{\beta}=(1-a^2)\frac{2a^2\cos^2{\tilde\alpha}}
{2a^2\cos^2{\tilde\alpha}-1},
\end{equation}
we should control the condition $0<\beta<\pi/2$, determining the region of variety
of parameters $(a,\alpha)$:
\begin{equation}
-\pi/4<\tilde\alpha<\pi/4,\quad
(\sqrt{2}\cos{\tilde\alpha})^{-1}<a<1,
\end{equation}
which is sketched in figure~4 below.

The last formulae define a domain, which we denote by ${\cal A}$, whose points
completely determine the geometry of octagon ${\cal F}$. Our aim is to
investigate the structure of ${\cal A}$.

To complete geometry description, the parameter $b$, pointed out in figure~1, is
calculated as
\begin{equation}\label{b}
b=(\sqrt{2}a\cos{\tilde\alpha})^{-1}.
\end{equation}

Note that the manifest dependence of the octagon parameters on the pair $(a,\alpha)$
is necessary in the different problems where geometry is not fixed. For instance,
$(a,\alpha)$ would be dynamical variables in topological field theory and gravity;
it is able to average over $(a,\alpha)$ in statistical physics, etc.

In order to define the Riemann surface $S$ based on the geometrical model, let us
recall that the opposite sides of ${\cal F}$ have the same lengths by construction.
We therefore have a uniquely defined isometry $g_k\in{\rm Isom}^+(\mathbb{D})$ mapping
geodesic boundary segment $s_{k+4}$ onto $s_k$ for all $k=\overline{0,3}$ (see figure~1,
right panel). For these isometries we get $g_k[{\cal F}]\cap{\cal F}=s_k$, where
$g_k[{\cal F}]$ means the set $\{g_k[z]|z\in{\cal F}\}$. Pasting sides
$s_{k+4}$ and $s_k$ together by identifying any $z\in s_{k+4}$ with $g_k[z]\in s_k$,
we obtain a closed surface of genus two that carries the hyperbolic metric
inherited from ${\cal F}$.

After introduction of four isometries $g_k$ and their inverses $g^{-1}_k$ generating
Fuchsian group $\Gamma$ with a single relation,
\begin{equation}
g_0 g^{-1}_1 g_2 g^{-1}_3 g^{-1}_0 g_1 g^{-1}_2 g_3={\rm id},
\end{equation}
it is purely to define surface $S$ as a quotient $\mathbb{D}/\Gamma$
for which $\pi: \mathbb{D}\rightarrow S$ is the natural covering map.
This is a Fuchsian model $\Gamma$ of the Riemann surface $S$.

In general, we can equip $S$ with a structure $\{U_p,f_p\}_{p\in S}$
by specifying a chart $U_p$ around each point $p\in S$ and a homeomorphism $f_p$
identifying $U_p$ with any of the open sets in $\mathbb{D}$ covering $U_p$.

It turns out that the different octagons may lead to the same surface. For this
reason, we mark a surface by generators of $\Gamma$. Two marked
surfaces $(S,\Gamma)$ and $(S^\prime,\Gamma^\prime)$ are called marking
equivalent if there exist an isometry $\gamma: S\to S^\prime$ satisfying
$g_k^\prime=\gamma g_k\gamma^{-1}$ ($k=\overline{0,3}$). Then all marking
equivalent surfaces form a marking equivalence class $[S,\Gamma]$ representing
the Riemann surface $S$ together with a structure defined on it.

It is useful sometimes to mark a surface by selecting a curve system
$\Sigma$ of simple closed geodesics on it. Then the marking equivalence also
means an existence of isometry $\gamma:S\to S^\prime$ sending $\Sigma\to\Sigma^\prime$.
In this case an equivalence class is formed by a pair $[S,\Sigma]$.

The set of all marking equivalence classes of Riemann surfaces is called the Teichm\"uller
space and is simply denoted by ${\cal T}_g$ in the case of the closed and compact Riemann
surfaces of genus $g$. The definition of ${\cal T}_g$ depends in general on choice of a marking
of Riemann surfaces. In any case, the real dimension of ${\cal T}_g$ like vector space
equals $6g-6$ in accordance with the Riemann--Roch theorem. We immediately note that the
Riemann surfaces, constructed with geometrical constraints imposed above, result only in
the subset of the total ${\cal T}_2$ of dimension six. In this sense we call such a space
as reduced Teichm\"uller one.

Coming back to our model, octagon ${\cal F}$ is a fundamental domain of the Fuchsian
group $\Gamma$, elements of which are hyperbolic: $|\Tr\gamma|>2$ for all
$\gamma\in\Gamma$ and depend on the form of octagon. In order to find generators $g_k$,
it is reasonable to introduce the set of auxiliary variables: $\omega_0=\omega_+$,
$\omega_1=\omega_-$, $\omega_2=\rmi\omega_+$, $\omega_3=\rmi\omega_-$, $\omega_5=0$, where
\begin{equation}
\omega_\pm=\frac{b\rme^{\rmi\alpha}(1-a^2)+a\rme^{\rmi\pi(1\mp1)/4}(1-b^2)}{1-a^2b^2}
\end{equation}
are functions of parameters $(a,\alpha)$.

Let us also define the matrices $M_k=M(\omega_k)$, where
\begin{equation}\label{M}
M(\omega)=\frac{\rmi}{\sqrt{1-|\omega|^2}}\left(
\begin{array}{cc}
	1& -\omega\\
	\overline{\omega}& -1
\end{array}
\right).
\end{equation}

The generators of $\Gamma$ are directly expressed in these terms as
\begin{equation}
g_k=M_kM_5, \qquad k=\overline{0,3}.
\end{equation}
This form is general and can be applied for arbitrary admissible octagon~\cite{ABCKS}.

Our calculations give us the manifest dependence of $g_0$ and $g_1$ on $(a,\alpha)$:
\begin{eqnarray}
g_0&=&N(a,\tilde\alpha)\left(
\begin{array}{cc}
	a(1-\tan{\tilde\alpha})& (a^2-\tan{\tilde\alpha})+\rmi(1-a^2)\label{g0}\\
	(a^2-\tan{\tilde\alpha})-\rmi(1-a^2) & a(1-\tan{\tilde\alpha})
\end{array}
\right),\\
g_1&=&N(a,\tilde\alpha)\left(
\begin{array}{cc}
	a(1+\tan{\tilde\alpha})& (1-a^2)+\rmi(a^2+\tan{\tilde\alpha})\\
	(1-a^2)-\rmi(a^2+\tan{\tilde\alpha}) & a(1+\tan{\tilde\alpha})
\end{array}
\right),
\end{eqnarray}
here
$$
N(a,\tilde\alpha)=\frac{-\cos{\tilde\alpha}}{\sqrt{(1-a^2)(2a^2\cos^2{\tilde\alpha}-1)}},
\qquad {\tilde\alpha}=\alpha-\frac{\pi}{4}.
$$

The remaining generators are simply obtained by rotations:
\begin{equation}\label{gn}
g_{2,3}=R_\frac{\pi}{2}g_{0,1}R^{-1}_\frac{\pi}{2}, \quad
g^{-1}_k=R_{\pi}g_kR^{-1}_{\pi}, \quad
R_\varphi=\left(
\begin{array}{cc}
	\exp{\left(\rmi{\varphi}/{2}\right)}& 0\\
	0 & \exp{\left(-\rmi{\varphi}/{2}\right)}
\end{array}
\right).
\end{equation}

It is convenient sometimes to represent the generators of $\Gamma$ by half turns
as follows. Let $p_k$ be the mid-point of $k$-th side, $k=\overline{0,3}$. Since the
opposite sides of octagon have the same lengths, the
generators are then written as $g_k=H(p_k)$ (see \cite{ABCKS}), where
\begin{equation}\label{mat1}
H(p)=\frac{-1}{1-|p|^2}\left(
\begin{array}{cc}
	1+|p|^2& 2p\\
	2\overline{p}& 1+|p|^2
\end{array}
\right).
\end{equation}
The operation of matrices $H(p)$ consists of composition of the half turn (rotation
with angle $\pi$) of geodesic segment around the origin $z=0$ and the half turn
around point $p$.

Due to symmetry of our model, $p_0=p_+$, $p_1=p_-$, $p_2=\rmi p_+$, and $p_3=\rmi p_-$,
where
\begin{equation}
p_\pm=\frac{\omega_\pm}{1+\sqrt{1-|\omega_\pm|^2}}.
\end{equation}.

Having got the Fuchsian group, it is possible to build the octagonal lattice with
a given fundamental octagon: action of the group elements $\gamma\in\Gamma$ on
${\cal F}$ produces the daughter cells tiling unit disk,
$\mathbb{D}=\bigcup_{\gamma\in\Gamma}\gamma[{\cal F}]$.

\section{Fenchel--Nielsen parameters and symplectic form}
 
A hyperbolic Riemann surface of genus $g$ without boundary always contains a system of
$3g-3$ simple closed geodesics that are neither homotopic to each other nor homotopically
trivial. Regardless of which curve system we choose, the cut along these geodesics always
decomposes surface into $2g-2$ pairs of pants (three-holed spheres), playing a role of
natural building blocks for Riemann surfaces (for instance, see \cite{Bu}).

In the case at hand, surface $S$ constructed is two-holed torus which can be decomposed
into two pairs of pants by a system of three closed geodesics. Such a surgery permits
to calculate the global Fenchel--Nielsen (FN) parameters: lengths of these geodesics and
twists, needed for further investigation and defined as follows.

\begin{figure}
\begin{center}
\includegraphics[width=4.5cm]{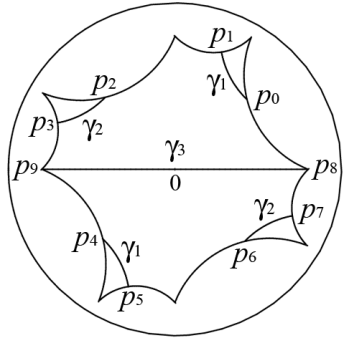}
\end{center}
\vspace*{-3mm}
\caption{\small Pants decomposition of the octagon with $a=0.8$, $\alpha=\pi/3$.}
\end{figure}

Let us consider the geodesic arcs from $p_0$ to $p_1$ and from $p_5$ to $p_4$ on the
octagon ${\cal F}$ (see figure~2). On the surface $S$ obtained by gluing the sides of
the octagon, these two arcs together form a smooth closed geodesic $\gamma_1$.
Similarly, a closed geodesic $\gamma_2$ is obtained from the arcs running from
$p_2$ to $p_3$ and from $p_7$ to $p_6$, respectively. The line $p_8p_9$ results in
a closed geodesic $\gamma_3$.
 
The triple $\gamma_1$, $\gamma_2$, $\gamma_3$ dissects $S$ into two pairs of
pants determined up to isometry by the hyperbolic lengths $\ell_k$, $k=\overline{1,3}$.
Immediate calculations yield
\begin{eqnarray}
\ell_{1,2}&\equiv&2~{\rm dist}_\mathbb{D}(p_+,p_-)=2~{\rm arccosh}~{\frac{a^2}{1-a^2}},\\
\ell_3&\equiv&2~{\rm dist}_\mathbb{D}(0,a)=2\ln{\frac{1+a}{1-a}},
\end{eqnarray}
where ${\rm dist}_\mathbb{D}(p_{n-1},p_n)={\rm dist}_\mathbb{D}(p_+,p_-)$ for
$n=1,3,5,7$.

When the pairs of pants are pasted together again to recover $S$, there arise additional
degrees of freedom at each $\gamma_k$, named the twist parameters $\tau_k$ and defined as
follows. On each pair of pants, one takes disjoint orthogonal geodesic arcs between each pair
of boundary geodesics. It is known that the feet of two perpendiculars on each geodesic are
diametrically opposite. Let us paste together two tubular neighborhoods of pair(s) of
pants with the boundaries of closed geodesics $\gamma^\prime_k$ and $\gamma^{\prime\prime}_k$ of
the same orientation and hyperbolic length and let us denote the weld by $\gamma_k$.
In principle, the feet of perpendiculars, arriving at the previously separated $\gamma^\prime_k$
and $\gamma^{\prime\prime}_k$, do not coincide on $\gamma_k$. The twist parameter $\tau_k$
is then the hyperbolic distance (shift) along $\gamma_k$ between
the feet of perpendiculars on opposite sides of the weld. Globally, the surfaces
arising from different $\tau_k$ are not in general isometric. This fact is often used for
investigation of Riemann surface deformations~\cite{Bu}-\cite{IT}.

Let us now concentrate on computational aspects.
One of convenient methods of geodesic lengths computation is the matrix formalism.
Here we follow the algorithms from \cite{ABCKS} based on it. 

We have already used the matrices $M_i$ ($i=\overline{0,3},5$) for finding the generators
of Fuchsian group in previous Section. However, in order to calculate the twists, it is
necessary to complete the set of generators by introducing matrix $M_4$, determined by
auxiliary parameter $\omega_4=2a/(1+a^2)$ (see (\ref{M})).

Introducing the set of parameters:
\begin{equation}\label{c}
c_1=-\frac{1}{2}\Tr(M_0M_1),\quad
c_2=-\frac{1}{2}\Tr(M_2M_3),\quad
c_3=-\frac{1}{2}\Tr(M_4M_5),
\end{equation}
\begin{eqnarray}
&d_1=\frac{1}{2}\Tr^2(M_0M_4M_5)-1,\quad
d_2=\frac{1}{2}\Tr^2(M_2M_1M_0)-1,&\nonumber\\
&d_3=\frac{1}{2}\Tr^2(M_5M_3M_2)-1,&
\end{eqnarray}
we can immediately check that $c_k=\cosh{(\ell_k/2)}$ ($k=\overline{1,3}$) and
\begin{equation}\label{d}
d_{1,2}+1=\frac{4}{(1-a^2)(1-b^2)},\quad
d_3+1=\frac{2}{(1-a^2)^2}.
\end{equation}

On the other hand, parameters $d_k$ are geometrically related
to the twists as
\begin{equation}\label{dt}
d_k=\frac{p}{c^2_k-1}(1+\cosh{\tau_k})-1,
\end{equation}
where $p=c^2_1+c^2_2+c^2_3+2c_1c_2c_3-1$.

Equating (\ref{d}) and (\ref{dt}), we obtain the twists in terms of model parameters:
\begin{equation}
\tau_{1,2}={\rm arccosh}\left[\frac{2a^2-1}{a^2(1-b^2)}-1\right],\quad
\tau_3=\ln{\frac{1+a}{1-a}}.
\end{equation}

It is known that the Teichm\"uller space of marked Riemann surfaces of genus
two forms a manifold homeomorphic to $\mathbb{R}^6$. This fact allows one
to identify the FN variables with global coordinates on it. However, the
Teichm\"uller space carries additional structure, namely, the Weil--Petersson (WP)
symplectic two-form. Actually, it is imaginary part of a natural K\"ahlerian metric.
Due to a theorem of Wolpert~\cite{Wo85,Wo82} (see also Thm.~3 in \cite{Wo08}),
WP two-form for compact closed Riemann surfaces of genus $g$ takes
on a particularly simple form in terms of FN variables,
\begin{equation}\label{wwp}
\omega_{\rm WP}=\frac{1}{2}\sum\limits_{k=1}^{3g-3}\rmd\ell_k\wedge\rmd\tau_k,
\end{equation}
with respect to any pants decomposition. It says in the sense of theoretical mechanics
that $\ell_k$ play the role of the action variables while $\theta_k=2\pi\tau_k/\ell_k$
are the angle variables. Indeed, the so-called simple Dehn twist $\theta_k\to\theta_k+2\pi$
gives us isometrically the same surface.

Using the pants decomposition presented in figure~2 and substituting the functions
$\ell_k$ and $\tau_k$ of $(a,\alpha)$ into (\ref{wwp}), the WP symplectic form becomes
\begin{equation}\label{wwp2}
\omega_{\rm WP}=\frac{8a}{(1-a^2)(2a^2\cos^2{\tilde\alpha}-1)}\rmd a\wedge \rmd\tilde\alpha,
\qquad \tilde\alpha=\alpha-\frac{\pi}{4}.
\end{equation}

To verify the uniqueness of the last formula, let us consider another pants
decomposition by changing connections between arc mid-points and main diagonal of
octagon, which give us new $\gamma^\prime_{1,2}$ and $\gamma^\prime_3$, respectively.
It is easily seen that a performed decomposition simply leads to the replacements,
\begin{equation}\label{z2}
a\leftrightarrow b,\qquad \tilde\alpha\leftrightarrow-\tilde\alpha,
\end{equation}
in the length and twist functions of the previous decomposition. We have
\begin{eqnarray}
\ell^\prime_{1,2}&=&2~{\rm dist}_\mathbb{D}(\rmi p_+,p_-)=2~{\rm arccosh}~{\frac{b^2}{1-b^2}},\\
\ell^\prime_3&=&2~{\rm dist}_\mathbb{D}(0,b)=2\ln{\frac{1+b}{1-b}},\\
\tau^\prime_{1,2}&=&{\rm arccosh}\left[\frac{2b^2-1}{b^2(1-a^2)}-1\right],\quad
\tau^\prime_3=\ln{\frac{1+b}{1-b}}.
\end{eqnarray}
where ${\rm dist}_\mathbb{D}(p_{n-1},p_n)={\rm dist}_\mathbb{D}(\rmi p_+,p_-)$ for
$n=2,4,6,8$.

Although we have obtained the set of new functions, the resulting two-form remains
the same, that is, $\omega^\prime_{\rm WP}=\omega_{\rm WP}$ due to the fact
${\rm sgn}\tau_k=-{\rm sgn}\tau^{\prime}_k$. Moreover, analysis shows
\begin{equation}
\rmd\ell_{1,2}\wedge\rmd\tau_{1,2}=\rmd\ell^\prime_{1,2}\wedge\rmd\tau^\prime_{1,2},\quad
\rmd\ell_3\wedge\rmd\tau_3=\rmd\ell^\prime_3\wedge\rmd\tau^\prime_3=0.
\end{equation}

Thus, we can conclude that i) the permission domain ${\cal A}$ of parameters
$(a,\alpha)$ is non-trivial symplectic manifold $({\cal A},\omega_{\rm WP})$;
ii) the Weil--Petersson symplectic two-form (\ref{wwp2}) is a closed and invariant
under action of the modular (sub)group $\mathbb{Z}_2$ represented by transformation
(\ref{z2}). Formally, we can treat the form (\ref{wwp2}) as an area element of
manifold ${\cal A}$, associated with the moduli space of Riemann surfaces under
consideration.

Furthermore, introducing the quantities,
\begin{eqnarray}
&&T_{1,2}\equiv\cosh{\frac{\tau_{1,2}}{2}}=\sqrt{\frac{2a^2-1}{2a^2(1-b^2)}},\quad
L_{1,2}\equiv\cosh{\frac{\ell_{1,2}}{2}}=\frac{a^2}{1-a^2},\\
&&T^\prime_{1,2}\equiv\cosh{\frac{\tau^\prime_{1,2}}{2}}=\sqrt{\frac{2b^2-1}{2b^2(1-a^2)}},\quad
L^\prime_{1,2}\equiv\cosh{\frac{\ell^\prime_{1,2}}{2}}=\frac{b^2}{1-b^2},
\end{eqnarray}
we can establish the following relations among them,
\begin{eqnarray}
&&L^{(\prime)}_3\equiv\cosh{\frac{\ell^{(\prime)}_3}{2}}=2L^{(\prime)}_1+1, \qquad
\tau^{(\prime)}_3=\ell^{(\prime)}_3/2,\\
&&L^\prime_1=T^2_1\frac{2L_1}{L_1-1}-1,\quad
T^\prime_1=\sqrt{\frac{L^2_1T^2_1+L_1T^2_1-L^2_1+1}{2L_1T^2_1-L_1+1}}.
\end{eqnarray}
These formulae reflect the symmetry of the model in terms of geometric constraints and
correspond to a special case of the surface with an order four automorphism, previously
studied in the literature (\cite{Sil}, Lm. 3.5) .

\section{Isoperimetric curves in ${\cal A}$}

We can also obtain additional information about structure of ${\cal A}$ by means of analysis
of principal geometric characteristics. One of those is an area, fixed by the Gauss--Bonnet
theorem and equal to $4\pi$ for genus two. Therefore, the area cannot obviously be a measure
of octagon deformation (evolution) preserving genus.

Simplest way to describe changes globally consists in consideration of a perimeter of
hyperbolic octagon. Within the present model, the perimeter is given by the formula:
\begin{equation}\label{Per}
P=8~{\rm arccosh}~\frac{1-a^2b^2+\sqrt{(1-a^2)^2+(1-b^2)^2}}{(1-a^2)(1-b^2)}.
\end{equation}

In a some sense, this characteristic is a good candidate due to invariance of $P$
under the octagon automorphisms and pants decomposition. It means that $P$ can take on
the same value for various values of $(a,\alpha)$. In this Section, we are aiming to
describe the corresponding orbits.

Evidence of this fact came from numerical investigation of information entropy
within the model of directed random walk on the Cayley tree generated by the Fuchsian
group (\ref{g0})-(\ref{gn}), where the perimeter and the entropy are connected~\cite{Naz}.
Moreover, the entropy had a maximum for octagonal lattice with $(a=2^{-1/4},\alpha=\pi/4)$,
what says about uniqueness of this configuration which should be proved here.

For further investigation, it is useful to introduce an auxiliary quantity,
\begin{equation}\label{EP}
E=2\left(\cosh{(P/8)}+1\right).
\end{equation}
Note that $E\to\exp{(P/8)}$ at $P\to\infty$.

For a given $E$, maximal and minimal values of parameter $a$ are found at $\tilde\alpha=0$
from equation
\begin{equation}\label{E(a)}
E=\frac{4a^2}{(1-a^2)(2a^2-1)}.
\end{equation}

We get
\begin{equation}\label{apm}
a_\pm(E)=\left(2\sqrt{E}\right)^{-1}\sqrt{3E-4\pm\sqrt{E^2-24E+16}}.
\end{equation}

It means that one can parametrize $a$ as follows
\begin{equation}\label{asol}
a(E,\varphi)=\left(2\sqrt{E}\right)^{-1}\sqrt{3E-4+\cos{\varphi}\sqrt{E^2-24E+16}},
\end{equation}
where periodic variable $\varphi\in[0,2\pi)$ is used.

Let us now solve algebraic equation
\begin{equation}
E^2-24E+16=0.
\end{equation}
We immediately obtain
\begin{equation}
E_{\rm reg}=12+8\sqrt{2},\quad
P_{\rm reg}=8~{\rm arccosh}~\left(5+4\sqrt{2}\right),\quad
a_{\rm reg}=2^{-1/4}.
\end{equation}
At $\tilde\alpha=0$, these quantities correspond to the regular hyperbolic octagon as
it must be. Thus, trajectory in ${\cal A}$ at $P_{\rm reg}$ is contracted to a point.
Moreover, $P_{\rm reg}$ is a minimal value of $P$ among possible ones. Therefore, the
maximal symmetry of the regular octagon explains an extremum of information entropy
observed previously. This fact could be important in description of the physical systems,
in which geometry carrier (two-holed torus) changes.

Substituting (\ref{asol}) in (\ref{Per}) and resolving the equation obtained with respect
to $\tilde\alpha$, we deduce that
\begin{equation}\label{alf}
\tilde\alpha(E,\varphi)=\arctan{\left[\left(\sqrt{2}E\right)^{-1}
\frac{\sqrt{(E-4)(E^2-24E+16)}\sin{\varphi}}{\sqrt{E-12-\cos{\varphi}\sqrt{E^2-24E+16}}}\right]}.
\end{equation} 

Equations (\ref{asol}), (\ref{alf}) allow us to see the orbits $P={\rm const}$, presented
in figure~3. The point corresponds to the parameters of the regular octagon
($P_{\rm reg}\approx24.45713$); cyclic curves are orbits for $P$ from $P=25$ to $P=41$ with step 2.

At $P\to\infty$, one obtains the asymptotics:
\begin{equation}
a_\infty(\varphi)=\frac{1}{2}\sqrt{3+\cos{\varphi}},\qquad
\tilde\alpha_\infty(\varphi)=\arctan{\frac{\sin{\varphi}}{\sqrt{2(1-\cos{\varphi})}}}.
\end{equation}  

\begin{figure}
\begin{center}
\includegraphics[width=8.5cm]{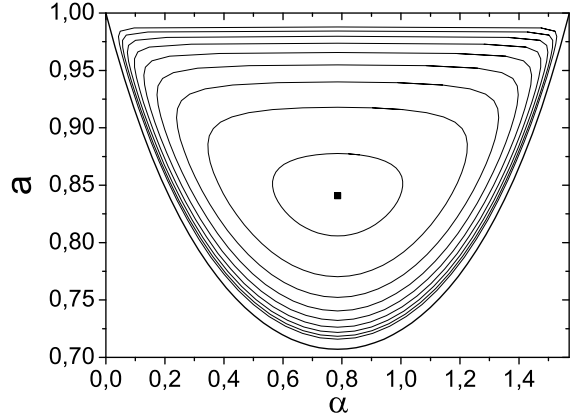}
\end{center}
\vspace*{-8mm}
\caption{\small Orbits of constant perimeter in the plane of octagon parameters.}
\end{figure}

We may assume that the functions $a(E,t)$, $\alpha(E,t)=\pi/4+\tilde\alpha(E,t)$ describe
some kind of octagon evolution, generated by translation operator $\partial/\partial t$
and determined by the initial coordinates $(a_0,\alpha_0)$ only. This diffeomorphism
produces continuous and smooth trajectory in ${\cal A}$ preserving the constant hyperbolic
length of the closed geodesics forming an octagon perimeter. Since the set of orbits is
dense in ${\cal A}$ there arises a possibility to geometrically quantize the symplectic
orbifold ${\cal A}$ in a spirit of \cite{Hurt}. In order to realise it, it is necessary to
consider a Weil-Petersson (WP) area ${\rm Area}(P_*)$ of the domain in ${\cal A}$, bounded
by isoperimetric orbit for some fixed $P_*$. Physically, ${\rm Area}(P_*)$ can be treated
as an action variable, that is, the only integral of motion
$\{a(E_*,t),\ \alpha(E_*,t)|t\in\mathbb{R}\}$,
where $E_*$ is related to $P_*$ by (\ref{EP}). Canonical quantization in terms of
${\rm Area}(P_*)$ and conjugate angle variable has to give us the number of quantum
states inside of the domain in ${\cal A}$. However, detailed development of quantum geometry
of ${\cal A}$ and the reduced Teichm\"uller space is a subject of another investigation which
will be published elsewhere. 

Here, using the WP symplectic form (\ref{wwp2}) and equations
(\ref{asol}), (\ref{alf}), we are limiting ourselves by introduction of WP area:
\begin{equation}
{\rm Area}(P_*)=\int_{P_{\rm reg}<P<P_*}
\frac{8a}{(1-a^2)(2a^2\cos^2{\tilde\alpha}-1)}\rmd a \rmd\tilde\alpha.
\end{equation}

This double integral is reduced to a single one:
\begin{equation}
{\rm Area}(P_*)=\int_{a_-(E_*)}^{a_+(E_*)}
\frac{16a}{(1-a^2)\sqrt{2a^2-1}}~{\rm arctanh}~f(E_*,a)\rmd a,
\end{equation}
where functions $a_\pm(E_*)$ are determined by (\ref{apm}), and
\begin{equation}
f(E_*,a)=\sqrt{\frac{(E_*-4)(1-a^2)}{E_*(1-a^2)-4}}\sqrt{1-\frac{E}{E_*}}
\end{equation}
is $\tan{\tilde\alpha}/\sqrt{2a^2-1}$ expressed in terms of $E_*$ and $a$;
$E$ is the function of $a$ given by (\ref{E(a)}).

Further calculations are performed numerically, and the result is demonstrated in figure~4.
Semi-analytical analysis shows that this curve can be approximated by a parabola,
$c_1(P-P_{\rm reg})^2+c_2(P-P_{\rm reg})$, with accuracy of the order $\Or(\exp{(-P/8)})$.
Best fit in the presented range of $P$ gives $c_1=0.05622$, $c_2=2.62132$.

\begin{figure}
\begin{center}
\includegraphics[width=8.5cm]{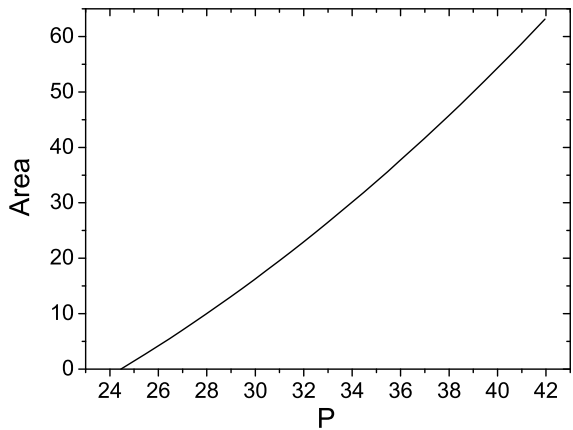}
\end{center}
\vspace*{-8mm}
\caption{\small WP area of domain bounded by isoperimetric curve $P={\rm const}$.}
\end{figure}

\section{Conclusions}

We studied the real-analytic structure of two-dimensional domain of geometric
parameters variety of hyperbolic octagon which is stable under an order four automorphism.
Carrying the Fenchel-Nielsen surgery and exploiting the Wolpert's theorem, it was shown
that this manifold is endowed with the fundamental symplectic two-form coming from the
Weil-Petersson (WP) geometry in the case of the closed compact surfaces in genus two. In
principal, the WP geometry also allows us to introduce the Riemannian (K\"ahlerian) metric.
However, this important problem remains unsolved and requires in general the use of
quasi conformal mappings~\cite{IT}, theory of which we have not touched in this paper.
Perspective of these investigations consists in the classical and quantum description
of free geometrodynamics of the surface in genus two. Attempts to do it have been already
performed in~\cite{NR,Lo}.

Additionally, the structure of parameter space has been considered from the point of view
of isoperimetric curves. The obtained trajectories might be treated as some kind of
hyperbolic octagon evolution and can be reformulated as quasi conformal mapping.
Furthermore, combining the formulae derived, we have evaluated the WP area of domain with
the boundary formed by the orbit of fixed perimeter. In principal, it gives a tool for
quantization of parameter space, associated with the moduli.



\begin{thebibliography}{99}

\bibitem{DHP}
D'Hocker~E and Phong~D~H 1988 {\it Rev. Mod. Phys.} {\bf 60} 917

\bibitem{Hurt}
Hurt~N~E 1983 {\it Geometric Quantization in Action: Applications of Harmonic
Analysis in Quantum Statistical Mechanics and Quantum Field Theory}
(D. Reidel Publishing Company)

\bibitem{Naz}
Nazarenko~A~V 2011 {\it Int. J. Mod. Phys.} {\bf B25} 3415 ({\it Preprint} math-ph/1112.2278)

\bibitem{Gutz}
Gutzwiller~M~C 1990 {\it Chaos in Classical and Quantum Mechanic} (New York: Springer-Verlag)

\bibitem{N}
Aurich~R, Sieber~M, and Steiner~F 1988 {\it Phys. Rev. Lett.} {\bf 61} 483

\bibitem{N1}
Ninnemann~H 1995 {\it Int. J. Mod. Phys.} {\bf B9} 1647

\bibitem{NR}
Nelson~J~E and Regge~T 1991 {\it Commun. Math. Phys.} {\bf 141} 211

\bibitem{Lo}
Loll~R 1995 {\it Class. Quant. Grav.} {\bf 12} 1655 ({\it Preprint} gr-qc/9408007)

\bibitem{ABCKS}
Aigon-Dupuy~A, Buser~P, Cibils~M, Kunzle~A~F, and Steiner~F 2005
{\it J. Math. Phys.} {\bf 46} 033513

\bibitem{Wo85}
Wolpert~S~A 1985 {\it Amer. J. Math.} {\bf 107(4)} 969

\bibitem{Sil}
Silhol~R 2007 {\it Comm. Math. Helv.} {\bf 82} 413

\bibitem{DNF}
Dubrovin~B~A, Novikov~S~P, and Fomenko~A~T 1984-1990 {\it Modern Geometry.
Methods and Applications} (Springer-Verlag) 

\bibitem{BS}
Buser~P and Silhol~R 2005 {\it Geometriae Dedicata} {\bf 115} 121

\bibitem{Bu}
Buser~P 1992 {\it Geometry and spectra of compact Riemann surfaces} (Birkh\"auser)

\bibitem{Wo81}
Wolpert~S 1981 {\it Comm. Math. Helv.} {\bf 56} 132

\bibitem{IT}
Imayoshi~Y and Taniguchi~M 1992 {\it An introduction to Teichm\"uller space} (Tokyo: Springer-Verlag)

\bibitem{Wo82}
Wolpert~S~A 1982 {\it Ann. of Math.} {\bf 115} 501

\bibitem{Wo08}
Wolpert~S~A 2008 The Weil-Petersson metric geometry {\it Preprint} math.DG/0801.0175v1

\end{thebibliography}
\end{document}